# Deep learning-based virtual refocusing of images using an engineered point-spread function

Xilin Yang,[†,‡,§] Luzhe Huang,[†,‡,§] Yilin Luo,[†] Yichen Wu,[†,‡,§] Hongda Wang,[†,‡,§] Yair Rivenson,[†,‡,§] and Aydogan Ozcan*,[†,‡,§,⊥],

[†]Electrical and Computer Engineering Department, [‡]Bioengineering Department, [§]California Nano Systems Institute (CNSI), [⊥]Department of Surgery, David Geffen School of Medicine, University of California, Los Angeles, California 90095, United States

**ABSTRACT:** We present a virtual image refocusing method over an extended depth of field (DOF) enabled by cascaded neural networks and a double-helix point-spread function (DH-PSF). This network model, referred to as W-Net, is composed of two cascaded generator and discriminator network pairs. The first generator network learns to virtually refocus an input image onto a user-defined plane, while the second generator learns to perform a cross-modality image transformation, improving the lateral resolution of the output image. Using this W-Net model with DH-PSF engineering, we extend the DOF of a fluorescence microscope by ∼20-fold. This approach can be applied to develop deep learning-enabled image reconstruction methods for localization microscopy techniques that utilize engineered PSFs to improve their imaging performance, including spatial resolution and volumetric imaging throughput.

Super-resolution imaging[1–6] and high-throughput volumetric[7–10] fluorescence microscopy provide unprecedented access to submicron-scale phenomena in various fields such as life sciences and engineering. However, improvements in the imaging resolution and throughput require relatively complex optical setups, usually through a time-consuming mechanical scanning procedure, which may also entail additional digital image registration and stitching procedures[10-12]. A particular method used for super-resolution imaging is based on localization microscopy[1-3,13]. Point spread function (PSF) engineering, including the use of astigmatic, multi-plane, double-helix (DH), and tetrapod PSFs, has been successfully used to improve the spatial resolution and depth of field (DOF) in localization microscopy [5,6,14-17]. However, the reconstruction of a sample's image that is convolved with an engineered PSF generally requires sparsity of the samples. Even for state-of-the-art localization algorithms[18], it is challenging to perform fast, accurate three-dimensional (3D) localization over an extended axial range with an increasing emitter density.

Recently emerging data-driven image reconstruction approaches have demonstrated performance advances for solving inverse problems in various microscopic imaging modalities[19-21]. Some of these methods help accelerate the 3D imaging process, preventing potential phototoxicity and photobleaching of the sample as well as improving the image resolution and throughput[22-31]. In particular, recent studies have demonstrated successful applications of deep learning methods for advancing 3D fluorescence microscopy. For example, Boyd *et al.* proposed DeepLoco[29] and a kernel-based loss function to outperform traditional 3D localization algorithms in terms of both speed and accuracy. Zhang *et al.* developed smNet[28], which can extract not only the 3D locations of fluorescence emitters but also their orientations and potential wave-front distortions. Nehme *et al.* demonstrated DeepSTORM3D[27], which utilizes a design method to jointly optimize the imaging PSF and the corresponding localization algorithm, extending the axial localization range up to 4 μm using a 1.45 NA/ 100 × objective lens. Wu *et al.* introduced Deep-Z[25], a deep learning-based virtual refocusing method that can refocus a given input image using a user-defined digital propagation matrix to an arbitrary surface within the sample volume, significantly improving the DOF and imaging throughput with only one input image. An extension of the same virtual refocusing approach using multiple input images has also been demonstrated for 3D volumetric imaging using recurrent neural networks[30].

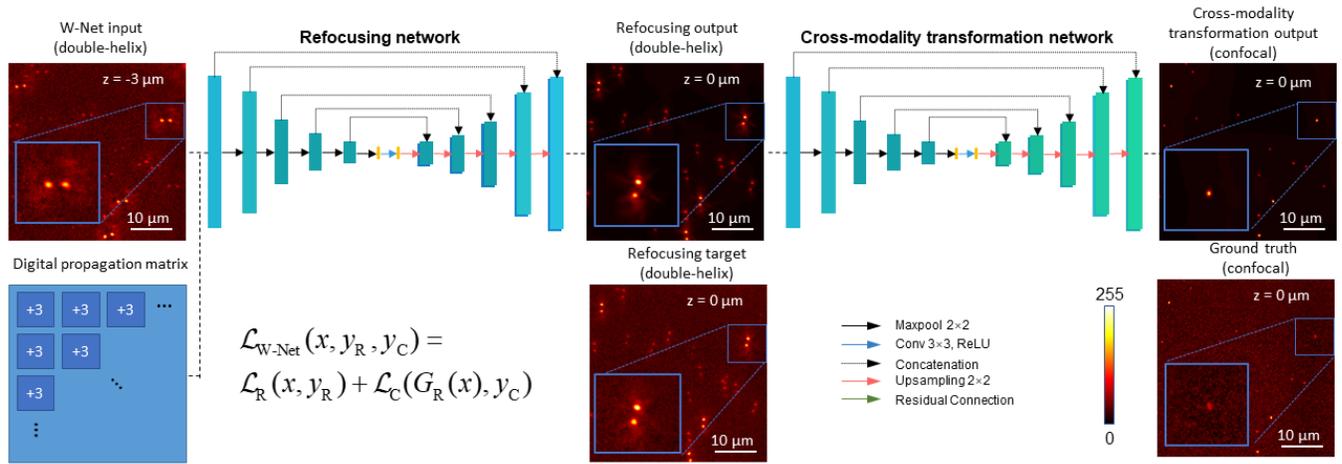

**Figure 1.** Structure of W-Net, containing two cascaded neural networks: (1) virtual image refocusing network and (2) cross-modality image transformation network optimized for DH-PSF. We use a joint training method for these two cascaded networks, where the output of the refocusing network is directly fed into the cross-modality image transformation network, and simultaneously minimize the losses of the refocusing ($L_R$) and cross-modality transformation ($L_C$) networks. $G_R$ is the generator of the refocusing network. The image intensity is log-scaled for a better contrast. Discriminators are not shown for simplicity (detailed in the Supporting Information.)

In this Letter, we present a deep learning-based method, referred to as W-Net (Figure 1), to perform both virtual refocusing and cross-modality image transformation of a single fluorescence microscopy image (input), acquired using an engineered PSF (i.e., double-helix PSF: DH-PSF[5]), onto user-defined planes within the sample volume. We trained our W-Net model as two cascaded neural networks to (1) virtually refocus a PSF engineered input image onto desired planes within the sample volume, and (2) perform image reconstruction at each virtually refocused plane, based on a cross-modality transformation method[24], which yields an output image equivalent to, e.g., a confocal fluorescence microscopy image of the same sample. The second step computationally resolves the spatial features of the sample convolved with the DH-PSF. Unlike standard iterative deconvolution techniques that can be used for images acquired with an engineered PSF, the presented method is based on a single pass-forward through a neural network and it does not require any iterations or mechanical scanning for 3D imaging of a sample owing to its digital refocusing capability.

Our W-Net design contains two cascaded U-Net structures[32], trained using a conditional generative adversarial network (cGAN)[33], as shown in Figure 1. Along with the DH-PSF input image, the first U-Net also receives, as a second input channel, a user-defined digital propagation matrix (DPM), which has the same size as the first channel. Each pixel value in the DPM determines the axial propagation distance of the corresponding pixel of the input image. Therefore, applying a series of DPMs on a single input image is equivalent to virtually scanning the specimen's volume.

In this study, we used experimentally acquired images for training and blindly testing W-Net in order to accurately capture and consider various complexities introduced by nonideal experimental conditions[19,24]. Using this new computational framework and DH-PSF, we digitally extended the DOF of the imaging system approximately 20 times, which was demonstrated by imaging nanoparticles. The presented method can be broadly applied for advancing localization microscopy techniques that utilize engineered PSFs by merging virtual refocusing with rapid volumetric image reconstruction.

To demonstrate the extended DOF and the DH-PSF reconstruction capability of our neural network model, we trained a W-Net model where the input images (50-nm fluorescent nanobeads) were acquired using the DH-PSF through a 63×/1.4-NA oil-immersion objective lens (see Methods section for microscopy system details); the native DOF of this objective lens is $\sim \pm$ 0.15 μm. In addition to this W-Net model that was trained with DH-PSF input images, we also trained a second W-Net model (for comparison) using wide-field images (as input) acquired on the same fluorescence microscope (63×/1.4-NA objective) by *removing* the DH-PSF phase mask. The ground truth image volumetric data corresponding to the same samples used for both W-Net models were acquired through a confocal microscope using the same objective lens (see Methods section). During the training phase of each W-Net model, we utilized digital propagation matrices such that both the input and target images were randomly defocused. At the blind testing stage, for quantifying the W-Net image inference, we designed the DPMs to virtually refocus the input images with different defocusing distances to a selected target plane (defined by z = 0 μm). We evaluated the quality of the W-Net output images refocused to z = 0 μm by calculating the image correlation coefficient between the output and ground truth (confocal) images over a region of 73.7 × 73.7 μm² using 75 individual nanobeads (Figure 2a). Furthermore, we utilized a customized localization algorithm, the Jaccard index (JI) and the lateral root mean square error (RMSE) metrics to quantify the localization performance of the W-Net output images, which also helped us measure the effective DOF (see the details in the Supporting Information). The results of these experimental analyses are summarized in Figures 2 and 3.

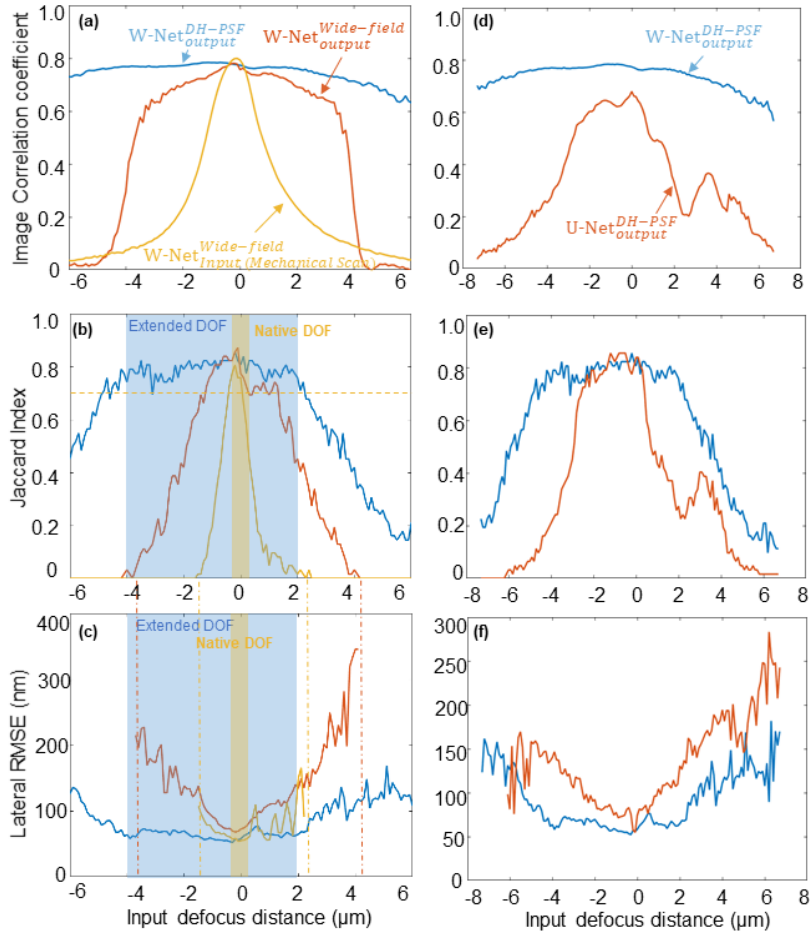

**Figure 2.** Quantifications of the (a–c) extended DOF by W-Net and DH-PSF engineering and (d–f) comparison to U-Net. The lateral RMSE is shown only when beads are detected (JI > 0, dot–dashed vertical lines in b, c in corresponding colors). In the left panels, the blue and red curves represent the W-Net outputs with DH and wide-field inputs, respectively. The yellow curves represent results of out-of-focus wide-field images obtained by mechanical scanning, which also correspond to the inputs used for the red line. The opaque yellow regions in b and c represent the native DOF defined by the objective lens, while the opaque blue region represents the extended DOF with the W-Net output. (d–f) Comparisons of W-Net to the single U-Net model. All metrics are calculated using confocal images as the ground truth over a 73.7 × 73.7 µm$^2$ region using 75 nanobeads.

We can directly compare the image correlation coefficients to qualitatively assess the extended DOF of the W-Net output (Figure 2a). Using input images that are captured with the DH-PSF, W-Net output significantly improves the imaging performance in the axial range of −6 to +6 µm. Without the use of DH-PSF, W-Net still successfully improves the image correlation coefficient, but the performance of this model deteriorates outside the range of ±3 µm. To quantify effective DOF we used localization analysis of the output images utilizing confocal images as reference (detailed in the Supporting Information); furthermore, we compared JI and the lateral RMSE of the localization results for detectability and localization accuracy, respectively (see Figure 2b,c). In Figure 2, the *horizontal* dashed line represents the average JI provided by the native DOF (orange highlighted region); the W-Net output can reach comparable results within an extended axial range (blue highlighted region), which is ~20 times larger compared with the native DOF of the objective lens. To further quantify the extended DOF, we set a series of JI thresholds with different tolerance levels and accordingly calculated the extended DOF and average lateral RMSE values, as shown in Table 1, which further confirm the success of W-Net output images.

| Tolerance | JI threshold | Average lateral RMSE (nm) | DOF (µm) |
|---|---|---|---|
| Native | 0.726 | 54.84 | 0.3 |
| 0% | 0.726 | 64.53 | 6.6 |
| 10% | 0.653 | 66.05 | 7.5 |
| 20% | 0.581 | 68.40 | 8.2 |

**Table 1**. Extended DOF and localization performance achieved by W-Net output images under different levels of tolerance for JI index degradation.

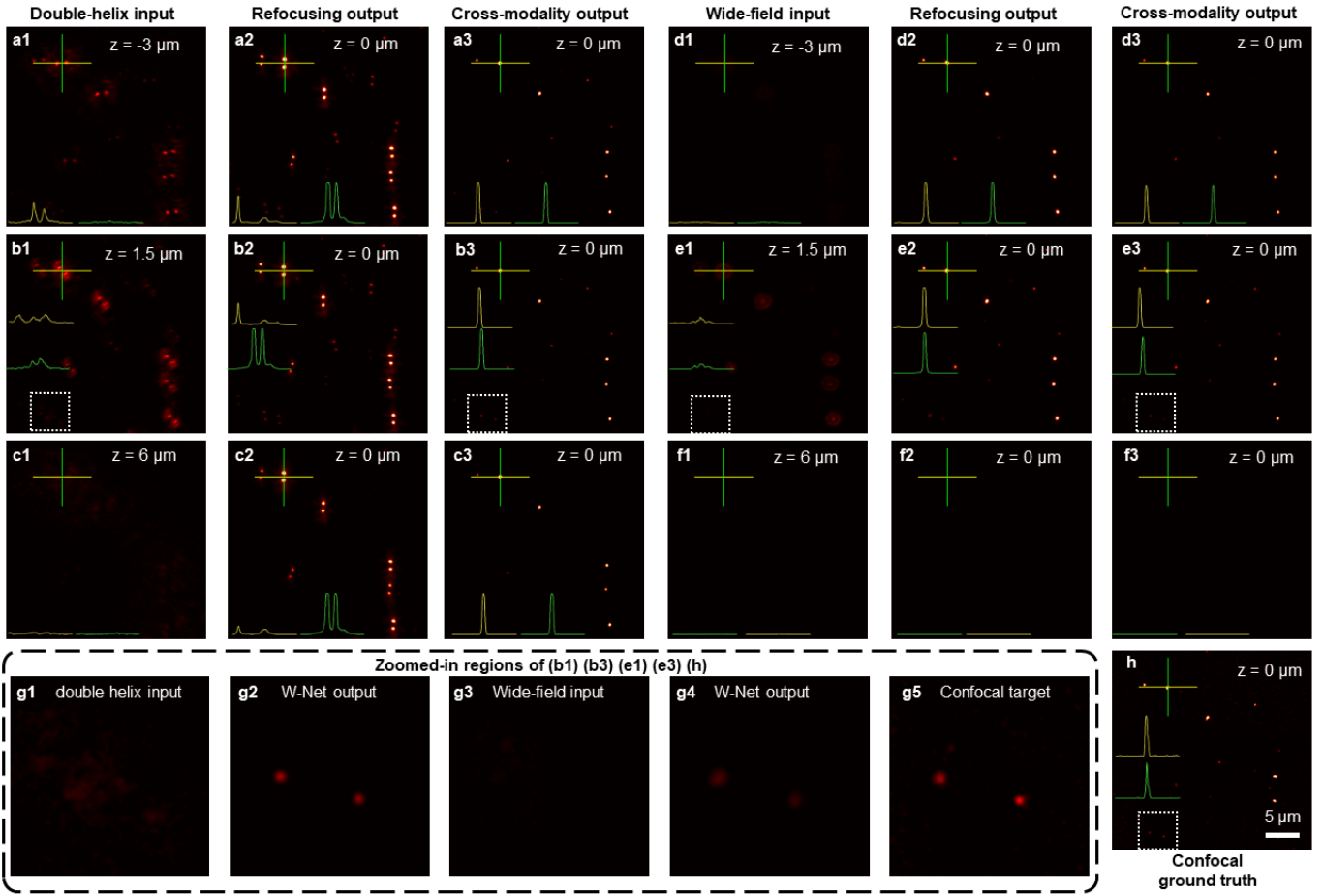

**Figure 3.** Test results obtained with the W-Net models for three different input planes at (a,d) $z$ = -3 μm, (b,e) $z$ = +1.5 μm, and (c,f) $z$ = +6 μm. (g1–5) Zoomed-in regions of (b1), (b3), (e1), (e3), and (h), respectively; the corresponding region is labeled with dashed squares. The first column shows the mechanically scanned images with DH-PSF, which were used as the input for W-Net. The second column shows the W-Net refocusing output. The third column shows the W-Net cross-modality output image of the same axial plane. The 4–6th columns (d–f) are similar but obtained with the conventional wide-field input images. (h) Ground truth obtained using confocal microscopy. The green and yellow lines are vertical and horizontal cross sections of a single nanobead labeled with the cross. The scale bar represents 5 μm for all images, except for the zoomed-in regions in (g).

In Figure 3, we further show the blind testing results obtained with W-Net models for three different input planes. At $z = -3$ μm (Figure 3a,d), the signal-to-noise ratio (SNR) of the scanned wide-field image is so low that no beads can be detected with the localization algorithm, which leads to a JI of 0. However, even when using such a defocused input, W-Net successfully refocuses a fraction of the nanobeads, scoring a JI of 0.17. With the use of DH-PSF, W-Net output further improves the JI to 0.77. The results also show an improvement in the lateral RMSE from 173.4 to 70.1 nm, which confirms that using DH-PSF with W-Net provides not only an extended DOF but also an improved localization accuracy. If the defocus distance is more significant, for example, $z = +6$ μm (Figure 2c,f), W-Net cannot retrieve adequate information from wide-field images, but still succeeds to collect a part of the DH-PSF input (JI=0.21) and reconstructs the image with a correlation coefficient of 0.63. Figure 5g (zoomed-in from b,e) presents an example of relatively weaker fluorescent beads at $z = +1.5$ μm, which are not visible in both of the input images, with the DH-PSF or wide-field. However, W-Net successfully reconstructs the two beads shown in this region. With the DH-PSF input, the resulting signal is stronger and gets closer to the confocal image target at $z = 0$ μm plane.

The W-Net structure has several advantages over a single U-Net structure performing both virtual refocusing and cross-modality image transformation[25]. Despite having approximately two-fold more number of parameters, allocating refocusing and cross-modality image transformation tasks to two separate networks eases the training burden. For example, when we applied the same learning rate of the W-Net model, U-Net could not converge on the same training dataset with the same axial range; furthermore, U-Net model requires more iterations in comparison to W-Net to achieve an acceptable performance despite the use of approximately 50% fewer parameters (refer to the Supporting Information for network structures). Overall, W-Net has a higher performance than that of U-Net when both are adequately trained using defocused DH-PSF images as the input and the confocal images as the target (ground truth) images. The latter is illustrated in Figure 2e,f: an over-fitting of the U-Net model occurs at around $z = +2\ \mu m$, which leads to a decrease in image quality. According to the three metrics that we used (i.e., lateral RMSE, JI, and image correlation coefficient), W-Net outperforms U-Net in DH-PSF reconstructions and provides a larger DOF with a superior lateral localization performance.

In conclusion, we demonstrated a virtual refocusing method enabled by cascaded neural networks (W-Net) and an engineered PSF, i.e., DH-PSF. This deep learning-enabled approach can be applied for image reconstruction in localization microscopy techniques that utilize engineered PSFs. This framework also paves the way for simultaneous super-resolution and high-throughput fluorescence imaging, owing to its cascaded refocusing and cross-modality image transformation network structure.

## METHODS

**Data acquisition and preprocessing.**

For nanobead data preparation, we imaged 50-nm fluorescent beads using an inverted scanning microscope (TCS SP8, Leica Microsystems) and 63 ×/1.4-NA objective lens (HC PL APO 63 ×/1.4-NA oil CS2, Leica Microsystems) using both wide-field and confocal imaging. We applied a SPINDLE system and transmissive DH-PSF phase mask (Double Helix Optics) in the wide-field microscope to acquire DH-PSF engineered images. We scanned along the x and y axes for 10 × 10 FOVs with an overlap of 10% for digital stitching, and scanned along the z axis with a step size of 0.1 μm to obtain 3D image data for each modality. These data were preprocessed with a triangular threshold[34] to normalize the data, subtract the background noise, and remove saturated pixels. We then stitched the FOVs with the ImageJ plugin MIST[35] and roughly registered the stitched confocal FOVs to the corresponding wide-field images with an ImageJ descriptor-based registration (https://imagej.net/Registration). For the training data, we cropped the whole FOV into 256 × 256-pixel regions with an overlap of 10%, while for testing, we used a 1024 × 1024-pixel region (512 × 512 cropped regions in Figure 3). The axial ranges for training were $\pm 8, \pm 4$, and $\pm 2$ μm for the DH-PSF, wide-field, and confocal cases, respectively.

To obtain a higher performance, we designed a dual-peak phase correlation method (DPPCM) in Matlab (MathWorks Inc.) for a finer image registration process with subpixel accuracy. This method calculates the normalized 2D cross-correlation matrix of the input and target images. It then finds the two local maxima of the correlation matrix, calculates their subpixel positions with a polynomial fitting within a 5 × 5 region. And the middle points of local maxima pairs are used to estimate the shift of the two imaging modalities. We further accelerated this algorithm, approximately 10 times, using a graphics processing unit (GPU); a period of ~0.15 s was required to process a 1024 × 1024 region image pair. The final training dataset contained ~110,000 image pairs for both wide-field to confocal and DH-PSF to confocal images. The testing image datasets were from different FOVs compared to the training images, and were not seen by the network before.

For details regarding the network structure, training configurations and the image reconstruction quality evaluation metrics, refer to the Supporting Information.

## ASSOCIATED CONTENT

Supporting Information available:
(1) Network structure and training configurations, and
(2) Reconstruction quality evaluation.

## AUTHOR INFORMATION

Corresponding Author
* E-mail: ozcan@ucla.edu

## ACKNOWLEDGMENT

The authors acknowledge Double Helix Optics (http://www.doublehelixoptics.com/) for providing the SPINDLE system.

## REFERENCES

(1) Betzig, E.; Patterson, G. H.; Sougrat, R.; Lindwasser, O. W.; Olenych, S.; Bonifacino, J. S.; Davidson, M. W.; Lippincott-Schwartz, J.; Hess, H. F. Imaging Intracellular Fluorescent Proteins at Nanometer Resolution. *Science* **2006**, *313* (5793), 1642–1645. https://doi.org/10.1126/science.1127344.

(2) Rust, M. J.; Bates, M.; Zhuang, X. Sub-Diffraction-Limit Imaging by Stochastic Optical Reconstruction Microscopy (STORM). *Nat. Methods* **2006**, *3* (10), 793–796. https://doi.org/10.1038/nmeth929.

(3) Hess, S. T.; Girirajan, T. P. K.; Mason, M. D. Ultra-High Resolution Imaging by Fluorescence Photoactivation Localization Microscopy. *Biophys. J.* **2006**, *91* (11), 4258–4272. https://doi.org/10.1529/biophysj.106.091116.

(4) Hell, S. W.; Wichmann, J. Breaking the Diffraction Resolution Limit by Stimulated Emission: Stimulated-Emission-Depletion Fluorescence Microscopy. *Opt. Lett.* **1994**, *19* (11), 780. https://doi.org/10.1364/OL.19.000780.

(5) Pavani, S. R. P.; Thompson, M. A.; Biteen, J. S.; Lord, S. J.; Liu, N.; Twieg, R. J.; Piestun, R.; Moerner, W. E. Three-Dimensional, Single-Molecule Fluorescence Imaging beyond the Diffraction Limit by Using a Double-Helix Point Spread Function. *Proc. Natl. Acad. Sci.* **2009**, *106* (9), 2995–2999. https://doi.org/10.1073/pnas.0900245106.

(6) Huang, B.; Wang, W.; Bates, M.; Zhuang, X. Three-Dimensional Super-Resolution Imaging by Stochastic Optical Reconstruction Microscopy. *Science* **2008**, *319* (5864), 810–813. https://doi.org/10.1126/science.1153529.

(7) Nguyen, J. P.; Shipley, F. B.; Linder, A. N.; Plummer, G. S.; Liu, M.; Setru, S. U.; Shaevitz, J. W.; Leifer, A. M. Whole-Brain Calcium Imaging with Cellular Resolution in Freely Behaving Caenorhabditis Elegans. *Proc. Natl. Acad. Sci.* **2016**, *113* (8), E1074–E1081.

(8) Abrahamsson, S.; Chen, J.; Hajj, B.; Stallinga, S.; Katsov, A. Y.; Wisniewski, J.; Mizuguchi, G.; Soule, P.; Mueller, F.; Darzacq, C. D. Fast Multicolor 3D Imaging Using Aberration-Corrected Multifocus Microscopy. *Nat. Methods* **2013**, *10* (1), 60.

(9) Schrödel, T.; Prevedel, R.; Aumayr, K.; Zimmer, M.; Vaziri, A. Brain-Wide 3D Imaging of Neuronal Activity in Caenorhabditis Elegans with Sculpted Light. *Nat. Methods* **2013**, *10* (10), 1013–1020. https://doi.org/10.1038/nmeth.2637.

(10) Fan, J.; Suo, J.; Wu, J.; Xie, H.; Shen, Y.; Chen, F.; Wang, G.; Cao, L.; Jin, G.; He, Q.; Li, T.; Luan, G.; Kong, L.; Zheng, Z.; Dai, Q. Video-Rate Imaging of Biological Dynamics at Centimetre Scale and Micrometre Resolution. *Nat. Photonics* **2019**. https://doi.org/10.1038/s41566-019-0474-7.

(11) Ku, T.; Swaney, J.; Park, J.-Y.; Albanese, A.; Murray, E.; Cho, J. H.; Park, Y.-G.; Mangena, V.; Chen, J.; Chung, K. Multiplexed and Scalable Super-Resolution Imaging of Three-Dimensional Protein Localization in Size-Adjustable Tissues. *Nat. Biotechnol.* **2016**, *34* (9), 973–981. https://doi.org/10.1038/nbt.3641.

(12) Winnubst, J.; Bas, E.; Ferreira, T. A.; Wu, Z.; Economo, M. N.; Edson, P.; Arthur, B. J.; Bruns, C.; Rokicki, K.; Schauder, D.; Olbris, D. J.; Murphy, S. D.; Ackerman, D. G.; Arshadi, C.; Baldwin, P.; Blake, R.; Elsayed, A.; Hasan, M.; Ramirez, D.; Dos Santos, B.; Weldon, M.; Zafar, A.; Dudman, J. T.; Gerfen, C. R.; Hantman, A. W.; Korff, W.; Sternson, S. M.; Spruston, N.; Svoboda, K.; Chandrashekar, J. Reconstruction of 1,000 Projection Neurons Reveals New Cell Types and Organization of Long-Range Connectivity in the Mouse Brain. *Cell* **2019**, *179* (1), 268-281.e13. https://doi.org/10.1016/j.cell.2019.07.042.

(13) Deschout, H.; Shivanandan, A.; Annibale, P.; Scarselli, M.; Radenovic, A. Progress in Quantitative Single-Molecule Localization Microscopy. *Histochem. Cell Biol.* **2014**, *142* (1), 5–17.

(14) Juette, M. F.; Gould, T. J.; Lessard, M. D.; Mlodzianoski, M. J.; Nagpure, B. S.; Bennett, B. T.; Hess, S. T.; Bewersdorf, J. Three-Dimensional Sub–100 Nm Resolution Fluorescence Microscopy of Thick Samples. *Nat. Methods* **2008**, *5* (6), 527–529. https://doi.org/10.1038/nmeth.1211.

(15) Shechtman, Y.; Sahl, S. J.; Backer, A. S.; Moerner, W. E. Optimal Point Spread Function Design for 3D Imaging. *Phys. Rev. Lett.* **2014**, *113* (13), 133902. https://doi.org/10.1103/PhysRevLett.113.133902.

(16) Holtzer, L.; Meckel, T.; Schmidt, T. Nanometric Three-Dimensional Tracking of Individual Quantum Dots in Cells. *Appl. Phys. Lett.* **2007**, *90* (5), 053902. https://doi.org/10.1063/1.2437066.

(17) Shechtman, Y.; Weiss, L. E.; Backer, A. S.; Sahl, S. J.; Moerner, W. E. Precise Three-Dimensional Scan-Free Multiple-Particle Tracking over Large Axial Ranges with Tetrapod Point Spread Functions. *Nano Lett.* **2015**, *15* (6), 4194–4199. https://doi.org/10.1021/acs.nanolett.5b01396.

(18) Sage, D.; Pham, T.-A.; Babcock, H.; Lukes, T.; Pengo, T.; Chao, J.; Velmurugan, R.; Herbert, A.; Agrawal, A.; Colabrese, S.; Wheeler, A.; Archetti, A.; Rieger, B.; Ober, R.; Hagen, G. M.; Sibarita, J.-B.; Ries, J.; Henriques, R.; Unser, M.; Holden, S. Super-Resolution Fight Club: Assessment of 2D and 3D Single-Molecule Localization Microscopy Software. *Nat. Methods* **2019**, *16* (5), 387–395. https://doi.org/10.1038/s41592-019-0364-4.

(19) Rivenson, Y.; Göröcs, Z.; Günaydin, H.; Zhang, Y.; Wang, H.; Ozcan, A. Deep Learning Microscopy. *Optica* **2017**, *4* (11), 1437. https://doi.org/10.1364/OPTICA.4.001437.

(20) de Haan, K.; Rivenson, Y.; Wu, Y.; Ozcan, A. Deep-Learning-Based Image Reconstruction and Enhancement in Optical Microscopy. *Proc. IEEE* **2020**, *108* (1), 30–50. https://doi.org/10.1109/JPROC.2019.2949575.

(21) Barbastathis, G.; Ozcan, A.; Situ, G. On the Use of Deep Learning for Computational Imaging. *Optica* **2019**, *6* (8), 921. https://doi.org/10.1364/OPTICA.6.000921.

(22) Belthangady, C.; Royer, L. A. Applications, Promises, and Pitfalls of Deep Learning for Fluorescence Image Reconstruction. *Nat. Methods* **2019**, *16* (12), 1215–1225. https://doi.org/10.1038/s41592-019-0458-z.

(23) Möckl, L.; Roy, A. R.; Moerner, W. E. Deep Learning in Single-Molecule Microscopy: Fundamentals, Caveats, and Recent Developments [Invited]. *Biomed. Opt. Express* **2020**, *11* (3), 1633. https://doi.org/10.1364/BOE.386361.

(24) Wang, H.; Rivenson, Y.; Jin, Y.; Wei, Z.; Gao, R.; Günaydın, H.; Bentolila, L. A.; Kural, C.; Ozcan, A. Deep Learning Enables Cross-Modality Super-Resolution in Fluorescence Microscopy. *Nat. Methods* **2019**, *16* (1), 103–110. https://doi.org/10.1038/s41592-018-0239-0.

(25) Wu, Y.; Rivenson, Y.; Wang, H.; Luo, Y.; Ben-David, E.; Bentolila, L. A.; Pritz, C.; Ozcan, A. Three-Dimensional Virtual Refocusing of Fluorescence Microscopy Images Using Deep Learning. *Nat. Methods* **2019**, *16* (12), 1323–1331. https://doi.org/10.1038/s41592-019-0622-5.

(26) Wu, Y.; Rivenson, Y.; Zhang, Y.; Wei, Z.; Günaydin, H.; Lin, X.; Ozcan, A. Extended Depth-of-Field in Holographic Imaging Using Deep-Learning-Based Autofocusing and Phase Recovery. *Optica* **2018**, *5* (6), 704. https://doi.org/10.1364/OPTICA.5.000704.

(27) Nehme, E.; Freedman, D.; Gordon, R.; Ferdman, B.; Weiss, L. E.; Alalouf, O.; Naor, T.; Orange, R.; Michaeli, T.; Shechtman, Y. DeepSTORM3D: Dense 3D Localization Microscopy and PSF Design by Deep Learning. *Nat. Methods* **2020**, *17* (7), 734–740. https://doi.org/10.1038/s41592-020-0853-5.

(28) Zhang, P.; Liu, S.; Chaurasia, A.; Ma, D.; Mlodzianoski, M. J.; Culurciello, E.; Huang, F. Analyzing Complex Single-Molecule Emission Patterns with Deep Learning. *Nat. Methods* **2018**, *15* (11), 913–916. https://doi.org/10.1038/s41592-018-0153-5.

(29) Boyd, N.; Jonas, E.; Babcock, H.; Recht, B. DeepLoco: Fast 3D Localization Microscopy Using Neural Networks. *bioRxiv* **2018**, 267096. https://doi.org/10.1101/267096.

(30) Huang, L.; Luo, Y.; Rivenson, Y.; Ozcan, A. Recurrent Neural Network-Based Volumetric Fluorescence Microscopy. *ArXiv Prepr. ArXiv201010781* **2020**.

(31) Luo, Y.; Huang, L.; Rivenson, Y.; Ozcan, A. Single-Shot Autofocusing of Microscopy Images Using Deep Learning. *ArXiv200309585 Phys.* **2020**.

(32) Ronneberger, O.; Fischer, P.; Brox, T. U-Net: Convolutional Networks for Biomedical Image Segmentation. *ArXiv150504597 Cs* **2015**.

(33) Mirza, M.; Osindero, S. Conditional Generative Adversarial Nets. *ArXiv14111784 Cs Stat* **2014**.

(34) Zack, G. W.; Rogers, W. E.; Latt, S. A. Automatic Measurement of Sister Chromatid Exchange Frequency. *J. Histochem. Cytochem.* **1977**, *25* (7), 741–753.

(35) Chalfoun, J.; Majurski, M.; Blattner, T.; Bhadriraju, K.; Keyrouz, W.; Bajcsy, P.; Brady, M. MIST: Accurate and Scalable Microscopy Image Stitching Tool with Stage Modeling and Error Minimization. *Sci. Rep.* **2017**, *7* (1), 4988. https://doi.org/10.1038/s41598-017-04567-y.